\documentclass[aps,prl,preprint,superscriptaddress,showpacs]{revtex4}

\usepackage{graphicx}

\begin{document}


\title{Giant anomalous Hall resistivity of the room temperature ferromagnet Fe$_{3}$Sn$_{2}$ \-- a frustrated metal with the {\it kagom\'{e}-bilayer} structure}

\author{T. Kida}
\email{kida@mag.cqst.osaka-u.ac.jp}
\affiliation{KYOKUGEN, Osaka University, 1-3 Machikaneyama, Toyonaka, Osaka 560-8531,
Japan}

\author{L. Fenner}
\affiliation{Department of Chemistry, University College London, 20 Gordon Street,
London, WC1H 0AJ, United Kingdom}

\author{ A. S. Wills}
\affiliation{Department of Chemistry, University College London, 20 Gordon Street,
London, WC1H 0AJ, United Kingdom}
\affiliation{The London Centre for Nanotechnolgy, 17-19 Gordon Street, London WC1H 0AH, United Kingdom}

\author{I. Terasaki}
\affiliation{Department of Applied Physics, Waseda University, 3-4-1 Okubo, Shinjuku,
Tokyo 169-8555, Japan}

\author{M. Hagiwara}
\affiliation{KYOKUGEN, Osaka University, 1-3 Machikaneyama, Toyonaka, Osaka 560-8531,
Japan}

\date{\today}

\begin{abstract}

We have investigated magnetic and transport properties of the {\it kagom\'{e}-bilayer} ferromagnet Fe$_{3}$Sn$_{2}$. A soft ferromagnetism and a large anomalous Hall effect are observed. The saturated Hall resistivity of Fe$_{3}$Sn$_{2}$ is 3.2~$\mu\Omega$cm at 300~K, which is almost 20 times higher than that of typical itinerant-ferromagnets such as Fe and Ni. The anomalous Hall coefficient $R_{{\rm s}}$ is 6.7$\times$10$^{-9}$~$\Omega$cm/G at 300~K, which is three orders of magnitude larger than that of pure Fe. 
$R_{{\rm s}}$ obeys an unconventional scaling to the longitudinal resistivity, $\rho_{xx}$, of $R_{{\rm s}} \propto \rho_{xx}^{3.3}$. Such a relationship cannot be explained by the skew and/or side-jump mechanisms and indicates that the origin of the anomalous Hall effect in  the frustrated magnet Fe$_{3}$Sn$_{2}$ is indeed extraordinary. 
\end{abstract}

\maketitle

\section{Introduction}
The Hall effect is the sideways deflection of an electron current due to an off-diagonal coupling 
between electric and magnetic field~\cite{Hurd}. 
In itinerant ferromagnets, such as Fe or Ni, the presence of a spontaneous magnetization allows 
the effect to occur in the absence of an external magnetic field leading to a giant Hall resistivity, 
termed the anomalous Hall effect (AHE). 
Understanding this phenomenon and its origins are amongst the most fundamental and long term problems 
in condensed matter physics~\cite{KL,Hurd,Nagaosa}. 

In itinerant ferromagnets the Hall resistivity ($\rho_{{\rm H}}$) consists of two contributions: 
\begin{eqnarray}
\rho_{{\rm H}}=-\rho_{xy}=R_{0}H_{{\rm ext}}+4\pi R_{1}M,\label{eq1}
\end{eqnarray}
where $R_{1}=(1-N_{{\rm d}})R_{0}+R_{{\rm s}}$, $R_{0}$ is the ordinary Hall coefficient, $R_{{\rm s}}$ 
the anomalous Hall coefficient, $H_{{\rm ext}}$ the external magnetic field, $M$ the magnetization, and 
$N_{{\rm d}}$ the demagnetizing factor. 
The first term of the right-hand side in Eq.~(\ref{eq1}) represents the ordinary Hall effect. 
The second term is in proportion to the magnetization and represents the AHE. 
The origins of the AHE are commonly separated into extrinsic and intrinsic mechanisms, though these were 
both first considered to be based on spin-orbit coupling and the spin polarization of conduction electrons. 
The extrinsic mechanisms are skew-scattering, where spin-orbit coupling breaks the left-right symmetry of 
the Mott scattering of charge carriers by an impurity~\cite{Smit}, and side-jump scattering, which 
corresponds to a lateral displacement of an electron perpendicular to the direction of its spin and 
wavevector~\cite{Berger}. 
An example of an intrinsic mechanism is that proposed by Karplus and Luttinger~\cite{KL} based on 
a ferromagnetic Hamiltonian with spin-orbit coupling. 
The eigenstates of such a system have the form of translationally invariant Bloch waves that interact 
through the interband matrix elements of the current operator, leading to an anomalous Hall current when 
the two spin states are unequal, $i.e.$ under the condition for ferromagnetism. 
Recently, alternative mechanisms for the intrinsic AHE in conventional ferromagnets has been suggested 
based on a Berry phase that arises as conduction electrons move through non-trivial spin 
configurations~\cite{Ye,Onoda,Jungwirth,Miyasato} and the orbital Aharonov-Bohm effect~\cite{Kontani1,Kontani2}, 
giving new directions for research into this extraordinary effect. 

Much interest was generated from the observation of the AHE in the pyrochlore ferromagnet 
Nd$_{2}$Mo$_{2}$O$_{7}$, as it was argued that the chiral magnetic ordering gave rise to a Berry phase 
in the electronic scattering~\cite{Tag01,Tag03}. 
Recent studies have, however, cast doubt on this explanation, as the changes to the AHE appear not 
to follow those to the magnetic structure induced by a magnetic field~\cite{Yas06,Sato07}. 
It therefore appears that other qualities are responsible for the AHE in Nd$_{2}$Mo$_{2}$O$_{7}$, 
presumably related to spin frustration \-- the competition between magnetic interactions. 
Its purest realization is in the geometrically frustrated magnets, where exchange geometries based on 
triangular plaquettes hinder the formation of conventional N\'{e}el order and allow the formation of 
a wide variety of exotic ground states, such as the quantum spin fluid~\cite{herbertsmithite,kapellasite}, 
topological spin glass~\cite{jarosites_1,jarosites_2,H3O_3,H3O_4,H3O_5}, 
and partially ordered spin structures~\cite{Gd2Ti2O7_1,Gd2Ti2O7_2}. 
The influences of spin frustration in frustrated ferromagnets is less well understood due to 
the scarcity of model systems. 
Despite this, the work on frustrated ferromagnets with localized spins, 
the spin ices~\cite{Spin_ice1,Spin_ice2,Spin_ice3,Spin_ice4,Spin_ice5} shows clearly the wealth of 
exotic behavior that can be hosted by frustrated ferromagnetic states~\cite{Monopole,Kasteleyn}. 
The challenge is now to develop an understanding of how geometric frustration enriches the physics that 
results from coupling the dual characters of localized spins and itinerant electrons, and effects 
such as the Kondo effect~\cite{Kondo} or the competition between itinerant and dipole 
anisotropies~\cite{Deakin,Richter,Inamdar}. 

The itinerant ferromagnet Fe$_{3}$Sn$_{2}$, where the Fe-ions make up a {\it kagom\'{e}-bilayer} structure, 
was recently proposed as a model system with which to explore the interplay between conduction electrons, 
localized spins and spin frustration~\cite{Fenner09}. 
While this intermetallic material has been known for some time~\cite{Nial,Trumpy,Malaman1,Malaman2,Caer1,Caer2}, 
its recent reexamination as a frustrated itinerant ferromagnet lead to the discovery of a re-entrant 
spin glass phase below $T_{g}\sim60$~K and the suggestion of a non-trivial canted ferromagnetic spin structure 
that rotates slowly from the {\it kagom\'{e}} planes towards the $c$-axis upon warming from $\sim60$~K 
to near the high temperature ferromagnetic transition $T_{{\rm c}}\simeq640$~K~\cite{Fenner09}. 
In this work, we have investigated the magnetotransport properties of Fe$_{3}$Sn$_{2}$ in the intermediate 
non-trivial ferromagnetic regime where the moments are rotated away from the $c$-axis and show 
it to possess an anomalously large $R_{{\rm s}}$ that cannot be understood in terms of the conventional 
AHE mechanisms. 
These results, suggest clearly that an alternative mechanism is responsible for the AHE in this frustrated 
magnet.

\section{Experimental}
Polycrystalline samples of Fe$_{3}$Sn$_{2}$ were prepared following the method given in Ref.~\onlinecite{Fenner09}. 
Laboratory X-ray diffraction analysis of the samples showed a small amount of FeSn and FeSn$_{2}$ impurity 
phases (5~\% at maximum). 
As these materials are antiferromagnets, they are not expected to have a significant effect on the measured 
data shown below. 
We prepared the sample for the transport measurements by cutting from a pressed pellet a monolith with dimensions 
of 4.0$\times$1.0$\times$0.5~mm$^{3}$. 
The magnetization measurements were performed by using a commercial SQUID magnetometer (MPMS-$XL7$, Quantum 
Design,~Inc.) in magnetic fields up to 7~T. 
The electrical resistivity was measured between 2 and 300 K by a conventional dc four-probe method with 
a current of 1 mA. 
The Hall resistance was measured between 100 and 300 K in magnetic fields up to 7~T. 
In the Hall resistance measurements, the Hall voltage $V_{{\rm H}}$ at each measured temperature was 
determined from the measured voltage 
$V(H_{{\rm ext}},I)$ as $V_{{\rm H}}(H_{{\rm ext}})=(1/4)(V(H_{{\rm ext}},I)-V(H_{{\rm ext}},-I)+V(-H_{{\rm ext}},-I)-V(-H_{{\rm ext}},I))$ 
so that unwanted signals from misalignments of the contact electrodes and from thermoelectric voltages 
were canceled.

\section{Results and Discussion}
Previous work has show this sample of Fe$_{3}$Sn$_{2}$ to display ferromagnetic ordering at $T_{{\rm c}}=640$~K, 
a slow rotation of Fe-moments from the $c$-axis towards the kagome plane on cooling down to 60 K, and 
a re-entrant spin glass phase below $T_{g}\sim60$~K~\cite{Fenner09}. 
The rotation of the moments from the $c$-axis is notable as it involves relaxation of the rules that 
the moments are collinear and equal, and instead allows the formation of a non-trivial ferromagnetic 
phase in which the moments are not collinear and differ in magnitude. 
This complexity of the magnetic order is expected to be partly responsible for the formation of 
the low temperature spin glass phase and may be the origin of the AHE presented in this study. 

\begin{figure}
\includegraphics[width=0.8\textwidth,keepaspectratio]{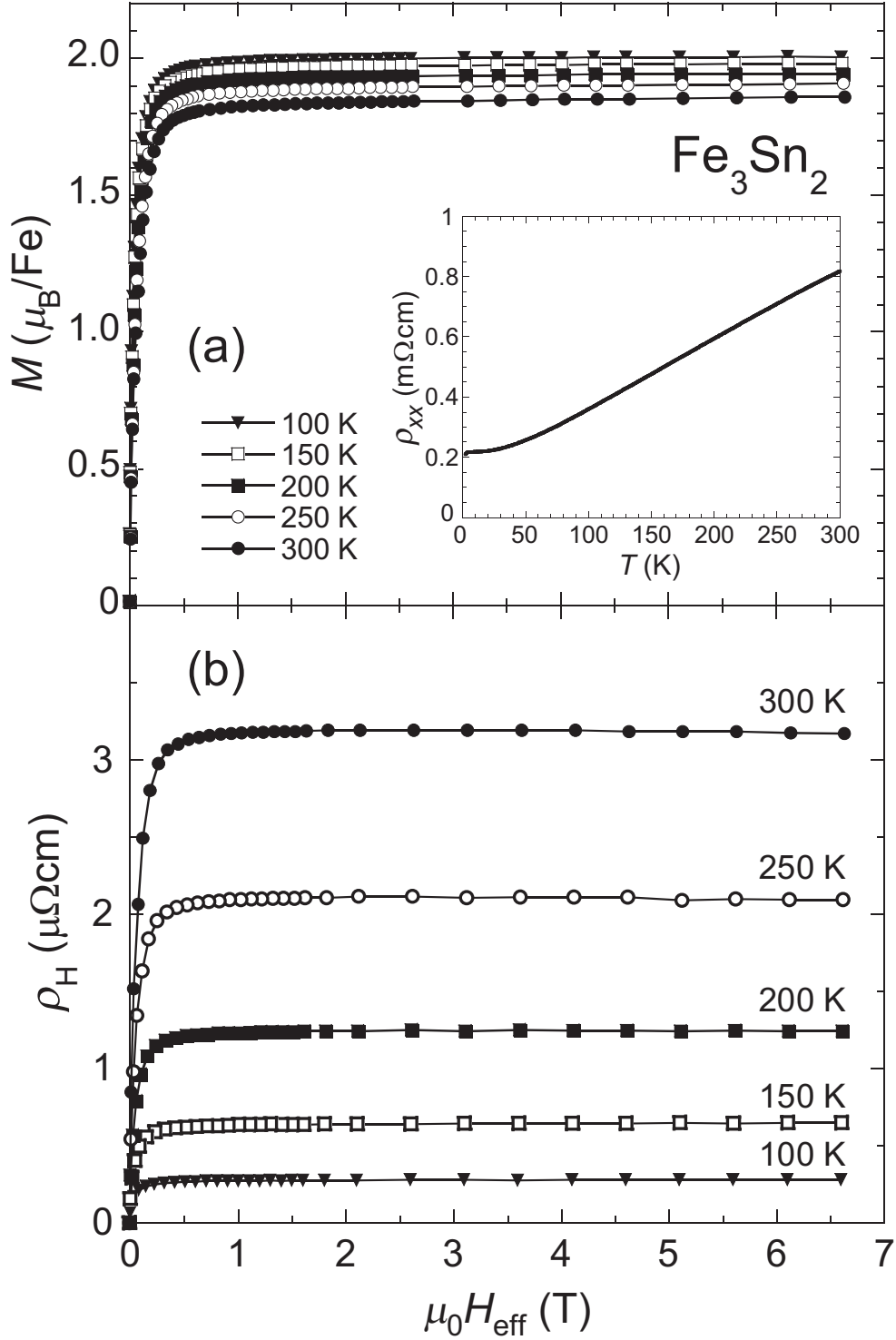} \caption{(a)~Magnetic field dependence of magnetization at various temperatures. The inset shows the temperature dependence of the electrical resistivity $\rho$. (b)~Magnetic field dependence of the Hall resistivities for a Fe$_{3}$Sn$_{2}$ polycrystalline sample at various temperatures.}
\label{Magnetization}
\end{figure}

The magnetization curves as a function of the effective magnetic field $H_{{\rm eff}}$ in the temperature range between 100 and 300~K are shown in Fig.~\ref{Magnetization}(a), where 
$H_{{\rm eff}}=H_{{\rm ext}}-4\pi N_{{\rm d}}M$. 
The value of the demagnetization correction $N_{{\rm d}}$(=~0.5667) is calculated following work on 
rectangular prisms by Chen \textit{et al}~\cite{Chen}. 
A rapid saturation and a weak remanent-magnetization are observed. 
The value of the saturation magnetization $M_{{\rm s}}$ estimated by the Arrott plot 
($M^{2}$ vs $H_{{\rm eff}}/M$) method is $2.0~\mu_{{\rm B}}/{\rm Fe}$ at 100~K, in good agreement with 
literature values~\cite{Fenner09,Caer1,Malaman2}. 

The electrical resistivity, $\rho_{xx}$, of Fe$_{3}$Sn$_{2}$ shows a metallic temperature dependence, 
and is 0.8~m$\Omega$cm at 300~K. 
The residual resistivity ratio (RRR) defined as $\rho_{xx}$(300~K)/$\rho_{xx}$(4~K), is close to 4. 
This value is much smaller than that of pure Fe and Ni~\cite{Volkenshtein} and indicates that 
the impurity and/or a grain-boundary scatterings are significant in this polycrystalline sample of Fe$_{3}$Sn$_{2}$.


The magnetic field dependence of $\rho_{{\rm H}}$ is shown in Fig.~\ref{Magnetization}(b) for Fe$_{3}$Sn$_{2}$ 
in the temperature range between 100 and 300~K. 
The shape of the $\rho_{{\rm H}}$-$H_{{\rm eff}}$ curves is very similar to that of the magnetization curves, $i.e.$, a rapid increase from $\mu_{0}H_{{\rm eff}}=$~0 to 0.5~T and a saturation in higher magnetic fields. 
The saturated value of $\rho_{{\rm H}}$ is 3.2~$\mu\Omega$cm at 300~K, that is almost 20 times higher than 
that of typical itinerant-ferromagnets Fe~\cite{Volkenshtein} and Ni~\cite{Kaul}, and is comparable to 
that of the pyrochlore molybdate Nd$_{2}$Mo$_{2}$O$_{7}$ at low temperatures~\cite{Yoshii}.

\begin{figure}
\includegraphics[width=0.9\textwidth,keepaspectratio]{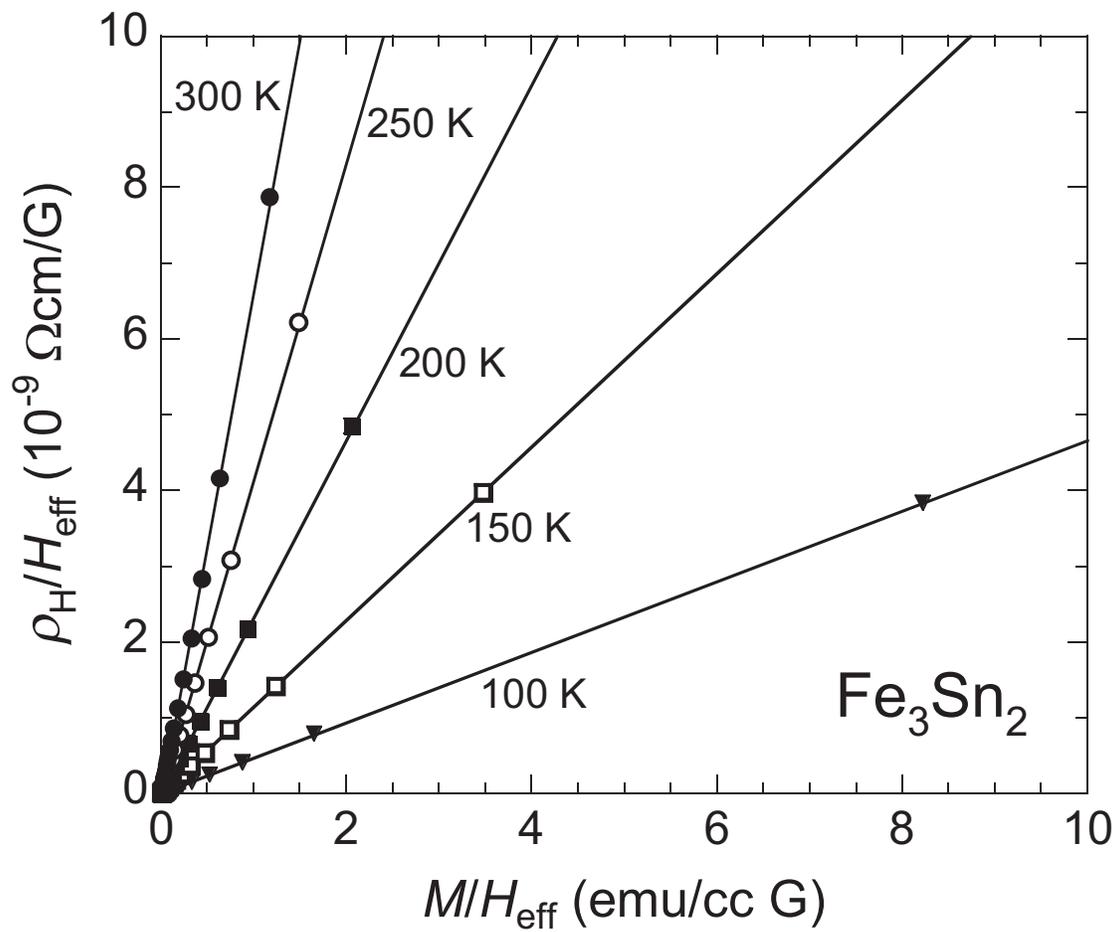}\caption{$\rho_{{\rm H}}/H_{{\rm eff}}$ vs $M/H_{{\rm eff}}$ at various
temperatures.}
\label{FieldDependence}
\end{figure}

Substituting $R_{1}[=(1-N_{{\rm d}})R_{0}+R_{{\rm s}}]$ into Eq.~(\ref{eq1}), 
and division of both sides by $H_{{\rm eff}}$, gives the following expression: 
\begin{eqnarray}
\rho_{{\rm H}}/H_{{\rm eff}}=R_{0}+4\pi(R_{0}+R_{{\rm s}})(M/H_{{\rm eff}}).\label{eq3} 
\end{eqnarray}
This relation allows the extraction of the values of the ordinary Hall coefficient, $R_{0}$, and 
the anomalous Hall coefficient, $R_{{\rm s}}$, from the intercept and slope of $\rho_{{\rm H}}/H_{{\rm eff}}$ 
plotted as a function of $M/H_{{\rm eff}}$, respectively. 
The values of $\rho_{{\rm H}}/H_{{\rm eff}}$ $vs.$ $M/H_{{\rm eff}}$ for Fe$_{3}$Sn$_{2}$ in the temperature 
range 100-300~K are shown in Fig.~\ref{FieldDependence}. 
$\rho_{{\rm H}}/H_{{\rm eff}}$ is highly linear in $M/H_{{\rm eff}}$, indicating that the AHE contribution 
of Fe$_{3}$Sn$_{2}$ is proportional to the magnetization. 
For all temperatures, the value of $|R_{0}|$ is of the order of 10$^{-12}$~$\Omega$cm/G. 
The carrier density, $n$, estimated by $R_{0}=|1/nec|$ ($e$ is the elementary electric charge; 
$c$ is the velocity of light.) is $10^{22}\sim10^{23}$~cm$^{-3}$, which is in the range of $n$ values 
of conventional metals including pure Fe and Ni~\cite{Volkenshtein}. 
In contrast, the value of $R_{{\rm s}}$ is 6.7$\times$10$^{-9}$~$\Omega$cm/G at 300~K, and is three orders of 
magnitude larger than $R_{{\rm s}}$ of pure Fe~\cite{Volkenshtein} and Ni~\cite{Kaul}.

\begin{figure}
\includegraphics[width=0.9\textwidth]{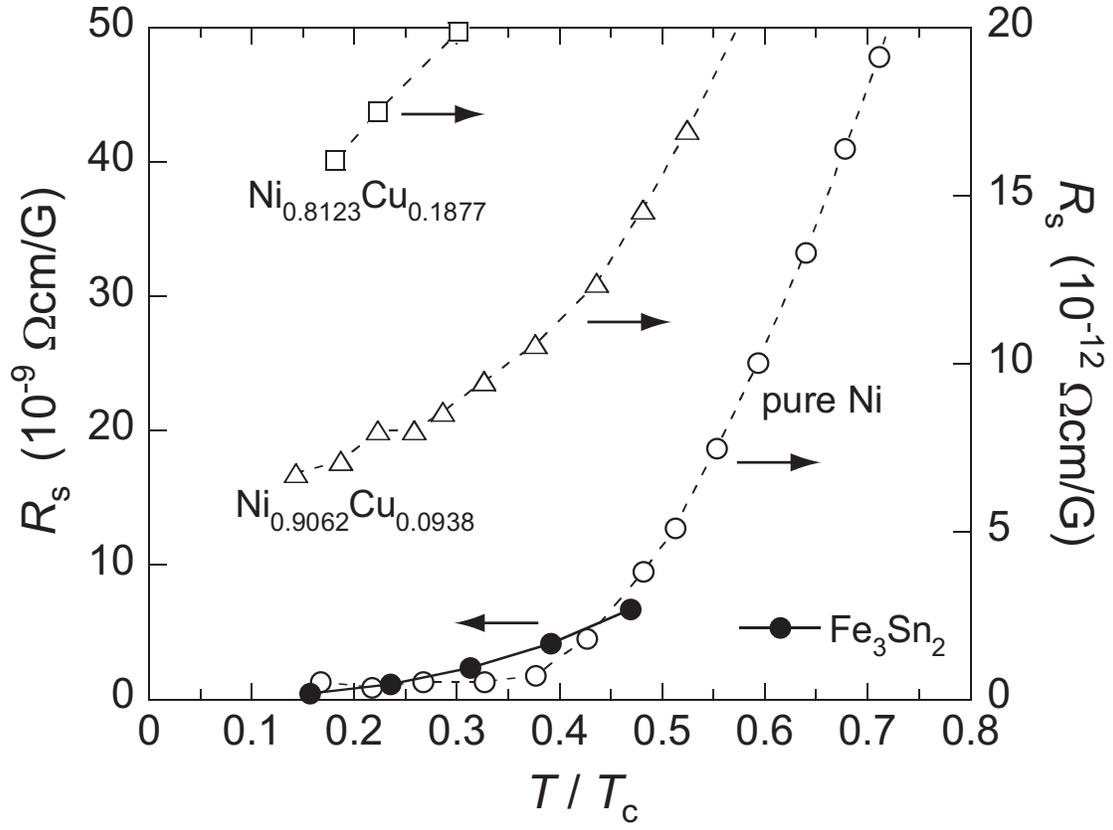} \caption{Anomalous Hall coefficient ($R_{{\rm s}}$) as a function of reduced temperature ($T/T_{{\rm C}}$) for Fe$_{3}$Sn$_{2}$, pure Ni, and Ni-Cu alloys.}
\label{AnomalousHallCoefficient}
\end{figure}

Figure~\ref{AnomalousHallCoefficient} shows $R_{{\rm s}}$ as a function of reduced temperature $T/T_{{\rm c}}$ 
for the present sample, pure Ni, and Ni-Cu alloys~\cite{Kaul}. 
For pure Ni, the $R_{{\rm s}}$ takes a maximum peak at around 0.9$T_{{\rm c}}$ 
(not shown in Fig.~\ref{AnomalousHallCoefficient}), and approaches towards zero below the peak~\cite{Kaul}. 
For the ferromagnetic Ni-Cu alloys, a residual component of $R_{{\rm s}}$ is observed at low $T/T_{{\rm c}}$ 
and increases with increasing Cu concentration. 
For example, the $R_{{\rm s}}$ of Ni$_{0.9062}$Cu$_{0.0938}$ is 7$\times$10$^{-12}$~$\Omega$cm/G at 
$T/T_{{\rm c}}=0.2$ and is almost 10 times higher than that of pure Ni. 
Although Fe$_{3}$Sn$_{2}$ shows nearly the same residual resistivity ratio as of Ni$_{0.9062}$Cu$_{0.0938}$, 
the residual component of $R_{{\rm s}}$ is small. 
The small $R_{{\rm s}}$ at low temperature indicates that the residual resistivity does not affect 
$R_{{\rm s}}$, and suggests that the origin of the large AHE on Fe$_{3}$Sn$_{2}$ cannot be explained by 
a simple extrinsic origin such as impurity effects.

\begin{figure}
\includegraphics[width=0.9\textwidth]{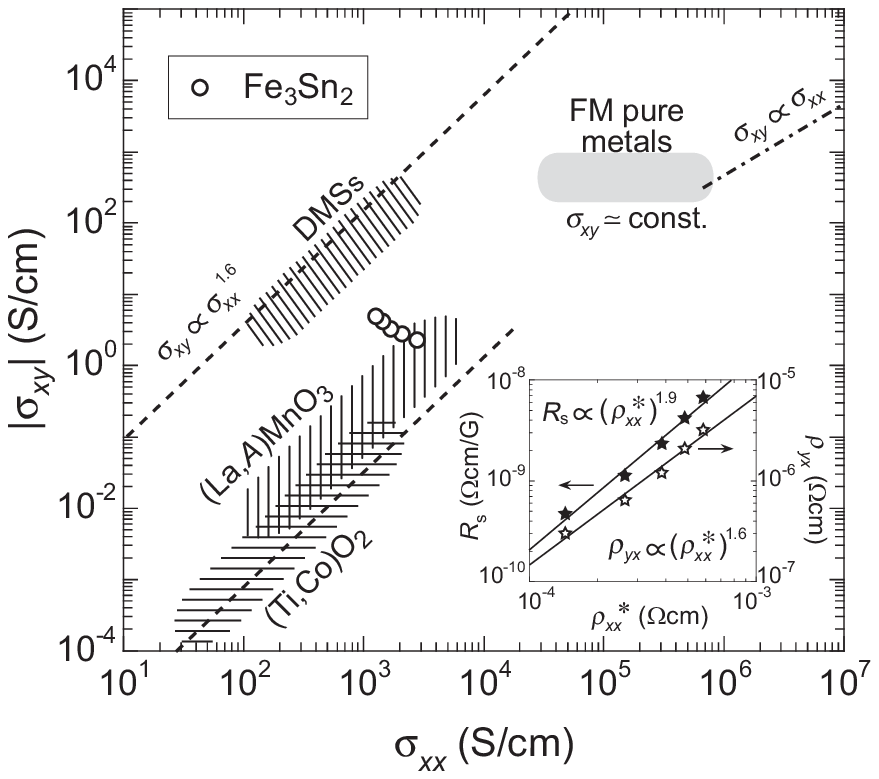} \caption{Absolute value of the anomalous Hall conductivity $|\sigma_{xy}|$ as a function of longitudinal conductivity $\sigma_{xx}$ in Fe$_{3}$Sn$_{2}$ (open circle), ferromagnetic (FM) pure metals, diluted magnetic semiconductors (DMSs), perovskite manganites ((La,~$A$)MnO$_{3}$), and Co-doped rutile TiO$_{2}$. The inset shows the $R_{{\rm s}}$ and $\rho_{yx}~(=\rho_{{\rm H}})$ as a function of $\rho_{xx}^{\ast}~(=\rho_{xx}-\rho_{0})$ for Fe$_{3}$Sn$_{2}$.}
\label{Map}
\end{figure}

Following to the preceding work by Miyasato $et~al$.~\cite{Miyasato}, we show in Fig.~\ref{Map} 
the absolute values of $|\sigma_{xy}|$ as a function of $\sigma_{xx}$ for the present sample (open circle) 
and for a range of other materials collated by Fukumura $et~al$.~\cite{Fukumura}, 
namely ferromagnetic (FM) pure metals, diluted magnetic semiconductors (DMSs), perovskite manganites 
((La,$A$)MnO$_{3}$), and Co-doped rutile TiO$_{2}$ on the same figure. 
Fe$_{3}$Sn$_{2}$ shows $\sigma_{xx}\sim10^{3}$~S/cm, and lies in the crossover region between metallic 
($10^{4}$~S/cm $<\sigma_{xx}<10^{6}$~S/cm) and hopping ($\sigma_{xx}<10^{4}$~S/cm) regimes. 
While many other ferromagnetic materials classified within the hopping regime are semiconducting, 
the temperature dependence of the resistivity shows Fe$_{3}$Sn$_{2}$ to display metallic conduction 
(inset of Fig.~\ref{Magnetization}a). 
Reflecting this extraordinary behavior, Fe$_{3}$Sn$_{2}$ does not obey the expected 
$\sigma_{xy}\propto\sigma_{xx}^{1.6}$ scaling (see open circles in Fig.~\ref{Map}).

Conventionally, the AHE has been characterized by the phenomenological relation 
between $R_{{\rm s}}$ and $\rho_{xx}$: 
\begin{eqnarray}
R_{{\rm s}}=a\rho_{xx}+b\rho_{xx}^{2},\label{eq4}
\end{eqnarray}
where $a$ and $b$ are constants. 
The linear and the quadratic terms in $\rho_{xx}$ are attributed to the skew-scattering~\cite{Smit} 
and the side-jump mechanism~\cite{Berger}, respectively. 
In many experimental cases of the AHE, a simpler relationship is found in which $R_{{\rm s}}$ is 
proportional to $\rho_{xx}^{n}$, with the index lying in the range $n=1\sim2$. 
In our case, the $R_{{\rm s}}$ of Fe$_{3}$Sn$_{2}$ does not obey Eq.~(\ref{eq4}) and is instead 
proportional to $\rho_{xx}^{3.3}$.


According to Eq.~(\ref{eq4}) a large $\rho_{xx}$, yields a large Hall resistivity. 
This scheme may provide a possible explanation for the large Hall resistivity of nanocomposite granular systems. 
As reported by Gao $et~al$.~\cite{Gao}, in co-sputtered Fe$_{x}$Sn$_{1-x}$ granular alloys a large saturated 
Hall resistivity is 8.68~$\mu\Omega$cm at room temperature and $\rho_{{\rm H}}\propto\rho_{xx}^{3.4}$. 
Clearly more detailed studies, such as annealing effect and impurity-doping effect are needed to understand 
the observed relationship between $R_{{\rm s}}$ and $\rho_{xx}$.

\section{Conclusions}
In summary, we have measured the magnetization, electrical resistivity and Hall resistivity of a 
{\it kagom\'{e}-bilayer} ferromagnet Fe$_{3}$Sn$_{2}$. 
A soft ferromagnetism and a large anomalous Hall effect are observed. 
The saturated Hall resistivity and the anomalous Hall coefficient of Fe$_{3}$Sn$_{2}$ are 3.2~$\mu\Omega$cm 
and 6.7$\times$10$^{-9}$~$\Omega$cm/G at 300~K, respectively. 
The $\sigma_{xy}\propto\sigma_{xx}^{1.6}$ scaling seen in many other AHE materials is not observed for 
Fe$_{3}$Sn$_{2}$ polycrystalline sample. 
Instead, the $R_{{\rm s}}$ is found to be proportional to $\rho_{xx}^{3.3}$, suggesting that the AHE on 
Fe$_{3}$Sn$_{2}$ is both unconventional and intrinsic, $i.e.$ it cannot be explained by the extrinsic 
effects of impurities. 
We speculate the origin of the effect is the non-trivial ferromagnetic structure and the frustration 
associated with the localized spins of the {\it kagom\'{e}-bilayer} structure. 

\section*{Acknowledgments}
This work was carried out under the foreign visiting researcher's program and the visiting researcher's 
program at KYOKUGEN in Osaka University. 
The authors thank Dr. K. Ogushi, Dr. K. Segawa, Prof. Y. Ando, Dr. W. Branford, and Prof. S. T. Bramwell 
for useful or fruitful discussion. 

This study was supported in part by Grant-in-Aid for Scientific Research (B) (No.~20340089) and 
for Scientific Research on Priority Areas "High Field Spin Science in 100T" 
(No.~451) from MEXT, Japan. 
ASW and LF would also like to thank the Royal Society and EPSRC (grant number EP/C534654) for funding.



\clearpage{} 

\section*{Figures}

%
\begin{figure}[h]
\includegraphics[width=0.9\textwidth,keepaspectratio]{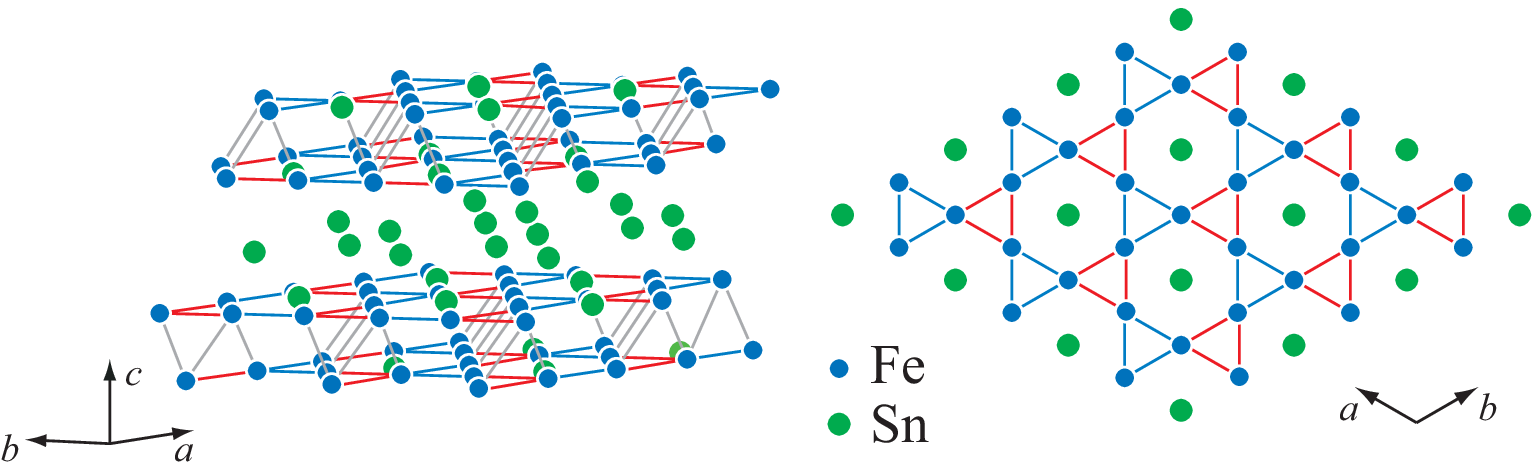}
\caption{The crystal structure of Fe$_{3}$Sn$_{2}$ is made up of Fe/Sn bilayers separated by Sn. The Fe ions form a {\it kagom\'{e}-bilayer} structure made up of 2 sizes of equilateral triangles shown in red and blue (color online)~\cite{Fenner09}.}
\label{Crystal_structure}
\end{figure}

\begin{thebibliography}{50}

\bibitem{Hurd}
C. M. Hurd, \textit{The Hall Effect in Metals and Alloys} (Plenum Press, New York, 1972). 

\bibitem{Nagaosa}
N. Nagaosa, J. Phys. Soc. Jpn. \textbf{75}, 042001 (2006). 

\bibitem{KL}
R. Karplus and J. M. Luttinger, Phys. Rev. \textbf{95}, 1154 (1954). 

\bibitem{Smit}
J. Smit, Physica \textbf{21}, 877 (1955); \textit{ibid}. \textbf{24}, 39 (1958). 

\bibitem{Berger}
L. Berger, Phys. Rev. B\textbf{2}, 4559 (1970). 

\bibitem{Ye}
J. Ye, Y. B. Kim, A. J. Millis, B. I. Shraiman, P. Majumdar, and Z. Te\v{s}anovi\'{c}, Phys. Rev. Lett. \textbf{83}, 3737 (1999). 

\bibitem{Onoda}
M. Onoda and N. Nagaosa, J. Phys. Soc. Jpn. \textbf{71}, 19 (2002). 

\bibitem{Jungwirth}
T. Jungwirth, Q. Niu, and A. H. MacDonald, Phys. Rev. Lett. \textbf{88}, 207208 (2002). 

\bibitem{Miyasato}
T. Miyasato, N. Abe, T. Fujii, A. Asamitsu, S. Onoda, Y. Onose, N. Nagaosa, and Y. Tokura, Phys. Rev. Lett. \textbf{99}, 086602 (2007). 

\bibitem{Kontani1}
H. Kontani, T. Tanaka, and K. Yamada, Phys. Rev. B\textbf{75}, 184416 (2007). 

\bibitem{Kontani2}
H. Kontani, T. Tanaka, D. S. Hirashima, K. Yamada, and J. Inoue, Phys. Rev. Lett. \textbf{100}, 096601 (2008). 

\bibitem{Tag01}
Y. Taguchi, Y. Oohara, H. Yoshizawa, N. Nagaosa, and Y. Tokura, Science \textbf{291}, 2573 (2001). 

\bibitem{Tag03}
Y. Taguchi, T. Sasaki, S. Awaji, Y. Iwasa, T. Tayama, T. Sakakibara, S. Iguchi, T. Ito, and Y. Tokura, Phys. Rev. Lett. \textbf{90}, 257202 (2003). 

\bibitem{Yas06}
Y. Yasui, T. Kageyama, T. Moyoshi, M. Soda, M. Sato, and K. Kakurai, J. Phys. Soc. Jpn. \textbf{75}, 084711 (2006).

\bibitem{Sato07}
M. Sato, J. Magn. Magn. Mater. \textbf{310}, 1021(2007).

\bibitem{herbertsmithite}
M. P. Shores, E. A. Nytko, B. M. Bartlett, and D. G. Nocera, J. Am. Chem. Soc. \textbf{127}, 13462 (2005).

\bibitem{kapellasite}
R. H. Colman, C. Ritter, and A. S. Wills, Chem. Mater. \textbf{20}, 6897 (2008).

\bibitem{jarosites_1}
A. S. Wills, Can. J. Phys. \textbf{79}, 1501 (2001).

\bibitem{jarosites_2}
A. S. Wills, A. Harrison, C. Ritter, R. I. Smith, Phys. Rev. B\textbf{61}, 6156 (2000).

\bibitem{H3O_3}
A. S. Wills, V. Depuis, E. Vincent, J. Hammann, R. Calemczuk, Phys. Rev. B\textbf{62}, R9264 (2000).

\bibitem{H3O_4}
A. S. Wills, G. S. Oakley, D. Visser, J. Frunzke, A. Harrison, and K. H. Andersen, Phys. Rev. B\textbf{64}, 094436 (2001).

\bibitem{H3O_5}
W. G. Bisson and A. S. Wills, J. Phys.: Condens. Matter. \textbf{20}, 452204 (2008).

\bibitem{Gd2Ti2O7_1}
J. D. M. Champion, A. S. Wills, T. Fennell, S. T. Bramwell, J. S. Gardner, and M. A. Green Phys. Rev. B\textbf{64}, 140407 (2001).

\bibitem{Gd2Ti2O7_2}
J. R. Stewart, G. Ehlers, A. S. Wills, S. T. Bramwell, and J. S. Gardner, J. Phys.: Condens. Matter. \textbf{16}, L321 (2004).

\bibitem{Spin_ice1}
S. T. Bramwell and M. J. Harris, J. Phys.: Condens. Matter {\bf 10} L215 (1998).

\bibitem{Spin_ice2}
M. J. Harris, S. T. Bramwell, and P. C. W. Holdsworth, Phys. Rev. Lett. \textbf{81}, 4496 (1998).

\bibitem{Spin_ice3}
R. F. Wang, C. Nisoli, R. S. Freitas, J. Li, W. McConville, B. J. Cooley, M. S. Lund, N. Samarth, C. Leighton, V. H. Crespi, and P. Schiffer, Nature \textbf{439}, 303 (2006).

\bibitem{Spin_ice4}
Y. Qi, T. Brintlinger, and J. Cumings, Phys. Rev. B\textbf{77}, 094418 (2008).

\bibitem{Spin_ice5}
A. S. Wills, R. Ballou, and C. Lacroix, Phys. Rev. B\textbf{66}, 144407 (2002).

\bibitem{Kasteleyn}
L. D. C. Jaubert, J. T. Chalker, P. C. W. Holdsworth, and R. Moessner, J. Phys.: Conf. Ser. \textbf{145}, 012024 (2009).

\bibitem{Monopole}
C. Castelnovo, R. Moessner, and S.L. Sondhi, Nature \textbf{451}, 42 (2008).

\bibitem{Kondo}
J. Kondo, Prog. Theoret. Phys. \textbf{32}, 37 (1964).

\bibitem{Deakin}
L. Deakin, M. J. Ferguson, A. Mar, J. E. Greedan, and A. S. Wills, Chem. Mater. \textbf{13}, 1407 (2001).

\bibitem{Richter}
M. Richter, J. Rusz, H. Rosner, K. Koepernik, I. Opahle, U. Nitzsche, and H. Eschrig, J. Magn. Magn. Mater. \textbf{272-276}, e251 (2004). 

\bibitem{Inamdar}
M. Inamdar, A. Thamizhavel, and S. Ramakrishnan, J. Alloys Compounds \textbf{480}, 28 (2009).

\bibitem{Fenner09} 
L. Fenner, A. A. Dee, and A. S. Wills, J. Phys.: Condens. Matter, \textit{submitted} (2009). 

\bibitem{Nial}
O. Nial, Svensk. Kern. Tidsk. \textbf{59}, 165 (1947).

\bibitem{Trumpy}
G. Trumpy, E. Both, C. Dj\'ega-Mariadassou, and P. Lecocq, Phys. Rev. B\textbf{2}, 3477 (1970).

\bibitem{Malaman1}
B. Malaman, B. Roques, A. Courtois, and J. Protas, Acta Cryst. B\textbf{32}, 1348 (1976). 

\bibitem{Malaman2}
B. Malaman, D. Fruchart, and G. L. Ca\"{e}r, J. Phys. F: Metal Phys. \textbf{8}, 2389 (1978). 

\bibitem{Caer1}
G. L. Ca\"{e}r, B. Malaman, and B. Roques, J. Phys. F: Metal Phys. \textbf{8}, 323 (1978). 

\bibitem{Caer2}
G. L. Ca\"{e}r, B. Malaman, L. H\"{a}ggstr\"{o}m, and T. Ericsson, J. Phys. F: Metal Phys. \textbf{9}, 1905 (1979). 

\bibitem{Chen}
D.-X. Chen, E. Pardo, and A. Sanchez, IEEE Trans. Magn. \textbf{38}, 1742 (2002). 

\bibitem{Volkenshtein}
N. V. Volkenshte\v{\i}n and G. V. Fedorov, Soviet Phys. JETP \textbf{11}, 48 (1960). 

\bibitem{Kaul}
S. N. Kaul, Phys. Rev. B\textbf{20}, 5122 (1979). 

\bibitem{Yoshii}
S. Yoshii, S. Iikubo, T. Kageyama, K. Oda, Y. Kondo, K. Murata, and M. Sato, J. Phys. Soc. Jpn. \textbf{69}, 3777 (2000). 

\bibitem{Fukumura}
T. Fukumura, H. Toyosaki, K. Ueno, M. Nakano, T. Yamasaki, and M. Kawasaki, Jpn. J. Appl. Phys. \textbf{46}, L642 (2007). 

\bibitem{Gao}
J. Gao, F. Wang, X. Jiang, G. Ni, F. Zhang, and Y. Du, J. Appl. Phys. \textbf{93}, 1851 (2003). 

\end{thebibliography}
\end{document}